\documentstyle[11pt]{article}

\newcommand{\be}{\begin{equation}}
\newcommand{\ee}{\end{equation}}
\newcommand{\bi}[1]{\vspace{-3mm} \bibitem{#1}}

\begin{document}

\begin{center}
{\large International Journal of Modern Physics B, 
Vol.20. No.3. (2006) pp.341-354}
\end{center}

\begin{center}
{\Large \bf Transport Equations from Liouville Equations for \\
Fractional Systems}
\vskip 5 mm

{\Large \bf Vasily E. Tarasov } \\
\vskip 3mm

{\it Skobeltsyn Institute of Nuclear Physics,
Moscow State University, Moscow 119992, Russia}
\vskip 5 mm

{E-mail: tarasov@theory.sinp.msu.ru}

\end{center}

\begin{abstract}
We consider dynamical systems that are described by fractional power of
coordinates and momenta.
The fractional powers can be considered
as a convenient way to describe systems in the fractional 
dimension space.  
For the usual space the fractional systems are non-Hamiltonian.
Generalized transport equation is derived from Liouville and
Bogoliubov equations for fractional systems.
Fractional generalization of average values and reduced distribution 
functions are defined.
Hydrodynamic equations for fractional systems 
are derived from the generalized transport equation.
\end{abstract}

PACS {05.40.-a, 02.50.Ey, 05.20.-y}


\section{Introduction}

Integrals and derivatives of fractional order \cite{OS,SKM}
have found many
applications in recent studies in physics \cite{Mai,WBG,Hilfer}.
The interest in fractional equations has been growing continually 
during the last few years because of numerous applications: 
kinetic theories of systems with chaotic and pseudochaotic 
dynamics \cite{Zaslavsky2,Zaslavsky8,Zaslavsky4,Zaslavsky9}; 
dynamics in a complex or porous media 
\cite{Nig,PLA05,PLA05-2,Physica,AP05,Chaos05}; 
electrodynamics \cite{CLZ,Plasma2005},
dynamical systems \cite{GM2,GM3,M,JPA2005-2,LMP,Stanislavsky}
and many others.

The new type of problems has rapidly increased interest in areas 
in which the fractal features of a process or the medium 
impose a necessity of applying the tools that are non-traditional  
in "regular" smooth physical equations. 
In many problems the real fractal structure of matter
can be disregarded and the medium can be replaced by
some smoothed continuous mathematical model where
fractional integrals appear \cite{PLA05,PLA05-2,AP05,Plasma2005}.
The order of fractional integral is equal
to the fractal mass dimension of medium and in this way one can 
take into account the fractality of media.
It was proved that integrals on net of fractals can be approximated 
by fractional integrals \cite{RLWQ} and that fractional integrals
can be considered as integrals over the space with fractional
dimension up to a numerical factor \cite{chaos,PRE05}.

It is known that Bogoliubov equations can be derived from the
Liouville equation and the definition of average value
Refs. \cite{Bog,Bog3,Gur,Petrina}.
In Refs. \cite{chaos}, the Liouville equation for fractional systems
is derived from the fractional normalization condition.
In \cite{PRE05}, the first Bogoliubov equation for fractional systems
is derived from the Liouville equation.
It is known that hydrodynamic equations can be derived from
generalized transport equation \cite{Enskog} which is directly derived from 
first Bogoliubov equation \cite{Bog,Gur,CC}.
In this paper, the hydrodynamic equations for fractional systems 
are derived from the generalized Enskog transport equations.

In Sec. 2, the fractional average values and some notations
are considered.
In Sec. 3, we define the reduced one-particle and two-particle
distribution functions.
In Sec. 4, the Liouville and first fractional Bogoliubov equations 
for fractional systems are considered.
In Sec. 5, we derive the fractional analog
of the Enskog transport equation. 
In Sec. 6, the hydrodynamic equations for fractional systems
are derived from the generalized transport equations.
Finally, a short conclusion is given in Sec. 7.

\section{Fractional Systems}

Let us consider a dynamical system that is described by the 
coordinates ${\bar q}_k$ and momenta ${\bar p}_k$ 
that satisfy the equations of motion:  
\be \label{bar1} \frac{d {\bar q}_k}{d {\bar t}}=\frac{{\bar p}_k}{M} ,
\quad
\frac{d {\bar q}_k}{d {\bar t}}=f_k({\bar q},{\bar p},{\bar t}) 
\quad k=1,...,n . \ee
Let us introduce the {\it dimensionless variables}
\[ q_k=\frac{{\bar q}_k}{q_0}, \quad p_k=\frac{{\bar p}_k}{p_0}, \quad
t=\frac{{\bar t}}{t_0}, \quad F_k=\frac{f_k}{F_0} , \]
where $q_0$ is a characteristic scale in the configuration space;
$p_0$ is a characteristic momentum, $F_0$ is a characteristic value
of force, and $t_0$ is a typical time.
Then Eq. (\ref{bar1}) has the form 
\be \label{bar2} \frac{d q_k}{d t}=\frac{p_k}{m} , \quad
\frac{d q_k}{d {\bar t}}=Z F_k(q,p,t) , \ee
where 
\be \label{A1} m=Mq_0/t_0p_0, \quad
Z=t_0 F_0 / p_0 . \ee
are dimensionless parameters. 
Using the dimensionless variables $(q,p,t)$, we can generalize Eq. (\ref{bar2}) 
for arbitrary powers of $q_k$ and $p_k$:
\be \label{bar3} \frac{d q^{\alpha}_k}{d t}=\frac{p^{\alpha}_k}{m} , 
\quad
\frac{d q^{\alpha}_k}{d {\bar t}}=Z F_k(q^{\alpha},p^{\alpha},t) , \ee
where  
\be \label{qa} q^{\alpha}_k =\beta(q) (q_k)^{\alpha}= sgn(q_k) |q_k|^{\alpha}, \ee
\be \label{pa} p^{\alpha}_k =\beta(p) (p_k)^{\alpha}= sgn(p_k) |p_k|^{\alpha}. \ee
Here $k=1,...,n$, and $\beta(x)=[sgn(x)]^{\alpha-1}$. 
The function $sgn(x)$ is equal to $+1$ for $x\ge0$, and $-1$ for $x<0$.

A system is called a {\it fractional system} if the phase space 
is described by the fractional powers of coordinates (\ref{qa}) 
and momenta (\ref{pa}). 
We can describe the fractional systems in the usual phase space $(q,p)$
and in the fractional phase space $(q^{\alpha},p^{\alpha})$.
In the second case, the equations of motion for the fractional 
systems are more simple. 
Therefore we use the fractional phase space.
The fractional space is considered as a space with the fractional measure 
that is used in the fractional integrals. 

The generalization of Hamiltonian system is described by 
\be \label{33} \frac{dq^{\alpha}_{k}}{dt}=
\frac{\partial H}{\partial p^{\alpha}_{k}},
\quad \frac{dp^{\alpha}_{k}}{dt}=
-\frac{\partial H}{\partial q^{\alpha}_{k}}, \ee
where $H$ is an analog of the Hamiltonian.
Using the fractional power Poisson brackets 
\be \label{PB} \{A,B\}_{\alpha}=
\sum^n_{k=1}\left(\frac{\partial A}{\partial q^{\alpha}_{k}}
\frac{\partial B}{\partial p^{\alpha}_{k}}-
\frac{\partial A}{\partial p^{\alpha}_{k}}
\frac{\partial B}{\partial q^{\alpha}_{k}} \right) , \ee
we rewrite Eq. (\ref{33}) in the from
\be \label{10n}
\frac{dq^{\alpha}_{k}}{dt}=\{q^{\alpha}_{k},H\}_{\alpha},
\quad \frac{dp^{\alpha}_{k}}{dt}=\{p^{\alpha}_{k},H\}_{\alpha}. \ee
These equations describe the system in the fractional phase space 
$(q^{\alpha},p^{\alpha})$.  
For the usual phase space $(q,p)$, Eq. (\ref{33}) has the form
\be \label{35} \frac{dq_{k}}{dt}=\frac{(q_kp_k)^{1-\alpha}}{\alpha^2}
\frac{\partial H}{\partial p_{k}},
\quad \frac{dp_{k}}{dt}=- \frac{(q_kp_k)^{1-\alpha}}{\alpha^2}
\frac{\partial H}{\partial q_{k}} , \ee
and these systems are non-Hamiltonian systems.
A classical system  
is called Hamiltonian if the right-hand sides of the equations
\be \label{qg-pf} \frac{dq_{k}}{dt}=g_k(q,p),
\quad \frac{dp_{k}}{dt}=f_k(q,p) \ee
satisfy the Helmholtz conditions \cite{Tar-tmf3}:
\be
\frac{\partial g_k}{\partial p_l}-\frac{\partial g_l}{\partial p_k}=0,
\quad
\frac{\partial g_k}{\partial q_l}-\frac{\partial f_l}{\partial p_k}=0,
\quad
\frac{\partial f_k}{\partial q_l}-\frac{\partial f_l}{\partial q_k}=0.
\ee
It is easy to prove these conditions are not satisfied
for Eq. (\ref{35}). 
Therefore the dymanical system (\ref{35}) is a non-Hamiltonian system.
The fractional phase space allows us to write Eq. (\ref{10n}) 
in the simple form (\ref{33}) and describe some non-Hamiltonian systems
as a Hamiltonian systems in generalized space.

If $dq^{\alpha}_k/dt=p^{\beta}_k/m$, then an analog of Hamiltonian 
\be \label{Hb} H_{\alpha,\beta}=
\sum^n_{k,l=1} \frac{\alpha p^{\alpha+\beta}_{k}}{m(\alpha + \beta)}
+U(q) . \ee 

The omega function for system (\ref{qg-pf}) is defined by 
\be \label{51}
\Omega=\sum^n_{k=1}\Bigl( \frac{\partial g_k}{\partial q_k}+
\frac{\partial f_k}{\partial p_k}\Bigr) ,
\ee
and describes the velocity of phase volume change.
If $\Omega<0$, then the system is called a dissipative system.
If $\Omega \not=0$, then the system is a generalized dissipative system.
For system (\ref{35}), 
the omega function (\ref{51}) is not equal to zero, and
the systems is the general dissipative system.

It is not hard to prove that Hamiltonian (\ref{Hb})
is connected with the non-Gaussian statistics.
Dissipative and non-Hamiltonian systems can have
the canonical Gibbs distribution as a solution of the stationary
Liouville equations  \cite{Tar-mplb}.
Using \cite{Tar-mplb}, it is easy to prove
that some of fractional systems can have
fractional Gibbs distribution 
\be \rho(q,p)=exp \ [{\cal F}-H_{\alpha, \beta}(q,p)]/kT, \ee
as a solution of Liouville equation 
for fractional systems \cite{chaos}.

\section{Fractional Average Values and Reduced Distributions}

\subsection{Fractional average values for configuration space}

Let us derive the fractional generalization of
average value of classical observable $A(q,p)$.
For configuration space, the usual average value is
\be <A>_1= \int^{+\infty}_{-\infty} A(x)\rho(x) dx , \ee
and can be written 
\be \label{If}
<A>_1=\int^y_{-\infty} A(x)\rho(x) dx +\int^{\infty}_y A(x)\rho(x) dx . \ee
Using 
\[ (I^{\alpha}_{+}f)(y)=
\frac{1}{\Gamma (\alpha)} \int^{y}_{-\infty}
\frac{f(x)dx}{(y-x)^{1-\alpha}} , \]
\[ \label{I-} (I^{\alpha}_{-}f)(y)=
\frac{1}{\Gamma (\alpha)} \int^{\infty}_{y}
\frac{f(x)dx}{(x-y)^{1-\alpha}} , \]
we rewtite Eq. (\ref{If}) in the form
\[ <A>_1=(I^{1}_{+}A\rho)(y)+(I^{1}_{-}A\rho)(y) . \]
The fractional generalization of this equation is 
\be \label{Aa} <A>_{\alpha}=
(I^{\alpha}_{+}A\rho)(y)+(I^{\alpha}_{-}A\rho)(y) . \ee
We can rewrite Eq. (\ref{Aa}) in the form
\be \label{FI5} <A>_{\alpha}= \frac{1}{2}
\int^{\infty}_{-\infty} [ (A\rho)(y-x)+ (A\rho)(y+x)] d\mu_{\alpha}(x) , \ee
where 
\be \label{dm}
d\mu_{\alpha}(x)=\frac{|x|^{\alpha-1} dx}{\Gamma(\alpha)}=
\frac{d x^{\alpha}}{\alpha \Gamma(\alpha)} , \quad
x^{\alpha}=sgn(x)|x|^{\alpha} . 
\ee
Equation (\ref{Aa}) defines the fractional generalization of the
average value for coordinate space.

\subsection{Fractional average values for phase space}

Let us introduce some notations to define the fractional
average value for phase space.
Tilde operators 
\[ T_{x_k} f(...,x_k,...)
=\frac{1}{2}\Bigl( f(...,x^{\prime}_k-x_k,...)
+f(...,x^{\prime}_k+x_k,...) \Bigr)  \]
allows us to rewrite
\[ \frac{1}{4}\Bigl(A(q'-q,p'-p,t)\rho(q'-q,p'-p,t) 
+A(q'+q,p'-p,t)\rho(q'+q,p'-p,t)+ \]
\[ +A(q'-q,p'+p,t)\rho(q'-q,p'+p,t) 
+A(q'+q,p'+p,t)\rho(q'+q,p'+p,t) \Bigr) \]
in the simple form
\[ T_q T_p (A(q,p,t)\rho(q,p,t)) . \]
For $k$ particle with
coordinates $q_{ks}$ and momenta $p_{ks}$, where $s=1,...,m$,
we define the operator  
\[ T[k]=T_{q_{k1}} T_{p_{k1}}...T_{q_{km}} T_{p_{km}} . \] 
For the $n$-particle system phase space, we use 
\[ T[1,...,n]=T[1]...T[n] . \] 

Let us define the integral operators $\hat I^{\alpha}_{x_k}$ by
\be \hat I^{\alpha}_{x_k} f(x_k)=
\int^{+\infty}_{-\infty}  f(x_k) d \mu_{\alpha} (x_k) , \ee
then Eq. (\ref{FI5}) has the form
\[ <A>_{\alpha}=\hat I^{\alpha}_{x} T_x A(x)\rho(x) . \]
For $k$-particle we use the operator 
\[ \hat I^{\alpha}[k]=
\hat I^{\alpha}_{q_{k1}} \hat I^{\alpha}_{p_{k1}} ...
\hat I^{\alpha}_{q_{km}} \hat I^{\alpha}_{p_{km}} , \] 
such that
\be \hat I^{\alpha}[k] f({\bf q}_k,{\bf p}_k)=
\int f({\bf q}_k,{\bf p}_k)
d \mu_{\alpha}({\bf q}_k,{\bf p}_k) , \ee
where $d \mu_{\alpha}({\bf q}_k,{\bf p}_k)$ is an elementary
$2m$-dimensional phase volume 
\[ d \mu_{\alpha}({\bf q}_k,{\bf p}_k)=(\alpha \Gamma(\alpha))^{-2m}
d q^{\alpha}_{k1}\wedge dp^{\alpha}_{k1} \wedge ... \wedge
d q^{\alpha}_{km}\wedge dp^{\alpha}_{km} . \]
For the $n$-particle system, we use 
\[ \hat I^{\alpha}[1,...,n]=\hat I^{\alpha}[1]...\hat I^{\alpha}[n] . \] 

Using the suggested notations, we can define
the fractional generalization of the average value for $n$-particle by 
\be <A>_{\alpha}=
\hat I^{\alpha}[1,...,n] T[1,...,n] A \rho_{n} .  \ee
In the simple case ($n=m=1$), we have
\be \label{AV2} <A>_{\alpha}=
\int^{\infty}_{-\infty} \int^{\infty}_{-\infty}
d\mu_{\alpha}(q,p) \ T_q T_p A(q,p)\rho(q,p) . \ee
The fractional generalization of normalization condition \cite{chaos}
can be written by
\[ <1>_{\alpha}=1 . \]

\subsection{Reduced distribution functions}

Let us consider a classical system with fixed number $n$ of
identical particles. 
Suppose $k$ particle is described by the {\it dimensionless} 
generalized coordinates $q_{ks}$ and generalized
momenta $p_{ks}$, where $s=1,...,m$.
We use the notations
${\bf q}_k=(q_{k1},...,q_{km})$ and
${\bf p}_k=(p_{k1},...,p_{km})$.
The state of this system is described by 
{\it dimensionless} n-particle distribution function $\rho_{n}$
in the $2mn$-dimensional phase space
\be \label{rh-t} \rho_{n}({\bf q},{\bf p},t)=
\rho({\bf q}_{1},{\bf p}_{1},...,{\bf q}_{n},{\bf p}_{n},t). \ee

We assume that function (\ref{rh-t}) is invariant under the
permutations of identical particles \cite{Bog2}:
\[ \rho(...,{\bf q}_{k},{\bf p}_{k},...,{\bf q}_{l},{\bf p}_{l},...,t)=
\rho(...,{\bf q}_{l},{\bf p}_{l},...,{\bf q}_{k},{\bf p}_{k},...,t) . \]
Then the average values can be simplified \cite{Bog2}.
Using the tilde distribution functions
\be \tilde \rho_n({\bf q},{\bf p},t)=T[1,...,n]\rho_n({\bf q},{\bf p},t) ,\ee
we define
\be \label{r1} \tilde \rho_{1}({\bf q},{\bf p},t)=
\tilde \rho({\bf q}_{1},{\bf p}_{1},t)=
\hat I^{\alpha}[2,...,n]\tilde \rho_{n}({\bf q},{\bf p},t) , \ee
which is one-particle reduced distribution function.
Obviously, that $\tilde \rho_1$ satisfies
the normalization condition \cite{chaos}:
\be \label{r3} 
\hat I^{\alpha}[1] \tilde \rho_{1}({\bf q},{\bf p},t)=1 . \ee
Two-particle reduced distribution function $\tilde \rho_2$
is defined by the fractional integration of 
$\tilde \rho_n$ over all ${\bf q}_{k}$ and ${\bf p}_{k}$,
except $k=1,2$:
\be \label{p2} \tilde \rho_{2}({\bf q},{\bf p},t)=
\tilde \rho({\bf q}_{1},{\bf p}_{1},{\bf q}_{2},{\bf p}_{2},t)=
\hat I^{\alpha}[3,...,n] \tilde \rho_{n}({\bf q},{\bf p},t) . \ee

\section{Liouville and Bogoliubov Equations for Fractional Systems}

Let us consider the Hamilton's equations for n-particle system in the form
\be \label{H3}
\frac{dq^{\alpha}_{ks}}{dt}=G^k_s(q^{\alpha},p^{\alpha}), \quad
\frac{dp^{\alpha}_{ks}}{dt}=Z F^k_s(q^{\alpha},p^{\alpha},t) , \ee
where $Z$ is defined in Eq. (\ref{A1}).
The evolution of $\rho_{n}$
is described by the Liouville equation \cite{chaos} for fractional system
\be \label{L1}
\frac{d \tilde \rho_n}{dt}+ \Omega_{\alpha} \tilde \rho_n =0. \ee
This equation can be derived \cite{chaos} from the fractional
normalization condition 
\be \hat I^{\alpha}[1,...,n] \tilde \rho_{n}({\bf q},{\bf p},t)=1 . \ee
In Eq. (\ref{L1}) the derivative $d/dt$ is a total time derivative
\[ \frac{d}{dt}=\frac{\partial}{\partial t}+
\sum^{n,m}_{k,s=1}\frac{dq_{ks}}{dt}\frac{\partial}{\partial q_{ks}}+
\sum^{n,m}_{k,s=1}\frac{dp_{ks}}{dt}\frac{\partial}{\partial p_{ks}}  \]
that can be written for the fractional powers 
\be \label{ttd3} \frac{d}{dt}=\frac{\partial}{\partial t}+
\sum^{n,m}_{k,s=1}
G^k_s\frac{\partial}{\partial q^{\alpha}_{ks}}+
Z\sum^{n,m}_{k,s=1}
F^k_s\frac{\partial}{\partial p^{\alpha}_{ks}} . \ee
The $\alpha$-omega function is 
\be  \label{o2} \Omega_{\alpha}=\sum^{n,m}_{k,s=1} \Bigl(
\{ G^k_s,p^{\alpha}_{ks}\}_{\alpha}+Z
\{q^{\alpha}_{ks},F^k_s\}_{\alpha} \Bigr) , \ee
where
\be \label{PB0} \{A,B\}_{\alpha}=\sum^{n,m}_{k,s=1}\left(
\frac{\partial A}{\partial q^{\alpha}_{ks}}
\frac{\partial B}{\partial p^{\alpha}_{ks}}-
\frac{\partial A}{\partial p^{\alpha}_{ks}}
\frac{\partial B}{\partial q^{\alpha}_{ks}} \right) . \ee
Using Eqs. (\ref{o2}) and (\ref{ttd3}), we get
Eq. (\ref{L1}) in the form
\be \label{r2}
\frac{\partial \tilde \rho_{n}}{\partial t}=\Lambda_{n} \tilde \rho_{n} , \ee
where $\Lambda_{n}$ is Liouville operator:
\be \label{Lam} \Lambda_{n} \tilde \rho_{n} =-
\sum^{n,m}_{k,s=1} \left(
\frac{\partial (G^k_s \tilde \rho_n)}{\partial q^{\alpha}_{ks}}+
Z\frac{\partial (F^k_s \tilde \rho_n)}{\partial p^{\alpha}_{ks}}  \right) . \ee

The Bogoliubov equations  \cite{Bog,Bog3,Gur,Petrina}
describe the evolution of the reduced distribution functions,
and  can be derived from the Liouville equation.
In Ref. \cite{PRE05}, we derive the first fractional 
Bogoliubov equation from Eq. (\ref{r2}) :
\be \label{er1-2} \frac{\partial \tilde \rho_{1}}{\partial t}+
\sum^m_{s=1}\frac{\partial (G^1_s \tilde \rho_{1})}{\partial q^{\alpha}_{1s}}
+Z\sum^m_{s=1}
\frac{\partial (F^{1e}_s\tilde \rho_{1})}{\partial p^{\alpha}_{1s}} =
(n-1)Z I(\tilde \rho_{2}). \ee
Here, $I(\tilde \rho_{2})$ is a term with two-particle
reduced distribution function, 
\be \label{I2} I(\tilde \rho_{2})=
- \sum^m_{s=1} \frac{\partial}{\partial p^{\alpha}_{1s}} 
\hat I^{\alpha}[2] F^{12}_s \tilde \rho_{2} . \ee
Equation (\ref{er1-2}) is called a 
{\it first Bogoliubov equation} for fractional systems. 

The physical meaning of the term $I(\tilde \rho_{2})$
is following:
The term $I(\tilde \rho_{2})d\mu_{\alpha}({\bf q},{\bf p})$
is a velocity of particle number change in $4m$-dimensional
elementary phase volume
$d\mu_{\alpha}({\bf q}_1,{\bf p}_2,{\bf q}_2,{\bf p}_2)$.
This change is caused by the interactions between particles.
If $\alpha=1$, then we have the first Bogoliubov equation for
non-Hamiltonian systems.

\section{Transport Equation for Fractional Systems}

Let us define the coordinate distribution 
(the density of number of particles) by the equation
\be  n({\bf q},t)= \hat I^{\alpha}[{\bf p}] 
\tilde \rho_1({\bf q},{\bf p},t) , \ee
where $\hat I[{\bf p}]$ is a fractional 
integration over the momenta
\be \hat I^{\alpha}[{\bf p}]=\prod^m_{s=1}\hat I^{\alpha}_{p_s}
=\hat I^{\alpha}_{p_1}...\hat I^{\alpha}_{p_m} . \ee
We can define the {\it local mean values} by
\be \langle A \rangle_{p,\alpha}=
\langle A \rangle_{p,\alpha}({\bf q},t) =
\frac{1}{ n({\bf q},t)}
\hat I^{\alpha}[{\bf p}] A({\bf q},{\bf p}) 
\tilde \rho_1({\bf q},{\bf p},t). \ee
In the general case, 
\be 
\langle A \rangle_{p,\alpha}\not=<A>_{\alpha} , 
\ee
and $\langle 1 \rangle_{p,\alpha}=1$. 
The fractional average value  $<A>_{\alpha}$ is connected with the
mean value  $\langle A \rangle_{p,\alpha}$ by 
\[ <A>_{\alpha}=\hat I^{\alpha}[{\bf q}]  n({\bf q},t)
\langle A \rangle_{p,\alpha}, \]
where $\hat I[{\bf q}]$ is a fractional 
integration over the coordinates ${\bf q}=(q_{1},...,q_{m})$.

Fractional analog of mean local velocity is 
\be V_s({\bf q},t)=\frac{1}{ n({\bf q},t)}
\hat I^{\alpha}[{\bf p}]
G_s({\bf q},{\bf p}) \rho_1({\bf q},{\bf p},t) , \ee
i.e.
\be V_s({\bf q},t)=\langle G_s \rangle_{p,\alpha} , \ee 
where $G_s=G^1_s({\bf q},{\bf p})$ is defined by (\ref{H3}).
We can consider $G_s$ 
as a fractional generalization of the velocity that has the form
\be \label{Gs} G_s=\frac{p^{\alpha}_s}{M} . \ee

For the fractional generalization of kinetic energy of relative motion 
\be 
\sum^m_{s=1} \frac{M}{2} (p^{\alpha}_s-V_s)^2 ,
\ee
we define density of this energy by
\be E({\bf q},t)=\frac{M}{2}
\hat I^{\alpha}[{\bf p}] 
({\bf G}-{\bf V})^2 \rho_1({\bf q},{\bf p},t) , \ee
where ${\bf G}={\bf p}^{\alpha}/m$, and ${\bf V}={\bf V}({\bf q},t)$. 
The local  temperature $T({\bf q},t)$ is defined 
by the mean kinetic energy of relative motion:
\[ T({\bf q},t)= \frac{2E({\bf q},t)}{3k_B n({\bf q},t)} . \]

To derive Enskog transport equation for fractional 
systems, we multiply both sides of (\ref{er1-2}) 
by the observable $A({\bf p})$, and integrate
with respect to momenta.
The first and second terms of left hand side of Eq. (\ref{er1-2})
are transformed by
\be \hat I[{\bf p}] A 
\frac{\partial \tilde \rho_1}{\partial t} =
\frac{\partial}{\partial t}
\hat I[{\bf p}] A \tilde \rho_1 =
\frac{\partial}{\partial t}  n({\bf q},t) 
\langle A \rangle_{p,\alpha}({\bf q},t), \ee

\be \hat I[{\bf p}] A 
\frac{\partial (G_s \tilde \rho_1)}{\partial q^{\alpha}_s} =
\frac{\partial}{\partial q^{\alpha}_s}
\hat I[{\bf p}] A G_s \tilde \rho_1 = 
\frac{\partial}{\partial q^{\alpha}_s}  n({\bf q},t) 
\langle A G_s \rangle_{p,\alpha} . \ee
Integrating by part the third term of Eq. (\ref{er1-2})
and using the boundary condition
\be \label{bound}
\lim_{p \rightarrow \pm \infty} \tilde \rho_1({\bf q},{\bf p},t) =0, 
\ee
we get
\[ \hat I[{\bf p}] A 
\frac{\partial (F_s \tilde \rho_1)}{\partial p^{\alpha}_s} =
\hat I[{\bf p}] \frac{\partial}{\partial p^{\alpha}_s} 
A F_s \tilde \rho_1 -\hat I[{\bf p}] F_s \tilde \rho_1 
\frac{\partial A}{\partial p^{\alpha}_s}= \]
\be = \Bigl(A F_s \tilde \rho_1 \Bigr)^{+\infty}_{-\infty}
-\hat I[{\bf p}] F_s \tilde \rho_1 
\frac{\partial A}{\partial p^{\alpha}_s}= 
-  n({\bf q},t) \langle 
F_s \frac{\partial A}{\partial p^{\alpha}_s} 
\rangle_{p,\alpha} . \ee
Then we use the usual assumption
\be \label{ass} 
\hat I[{\bf p}] A({\bf p}) I(\tilde \rho_{2})=0 . \ee
for 
$A=M$, $A=p^{\alpha}_s$ and $A ={\bf p}^{2\alpha}$.

Finally, we obtain the  
Enskog transport equation for fractional systems:
\be \label{EE} \frac{\partial}{\partial t} \left(  n({\bf q},t) 
\langle A \rangle_{p,\alpha}\right) +
\frac{\partial}{\partial {\bf q}^{\alpha}} \left( n({\bf q},t) 
\langle A {\bf G} \rangle_{p,\alpha} \right) = 
 n({\bf q},t) Z  \langle 
{\bf F} \frac{\partial A}{\partial {\bf p}^{\alpha}} 
\rangle_{p,\alpha} , \ee
where
\[ {\bf F} \frac{\partial }{\partial {\bf p}^{\alpha}} =\sum^m_{s=1}
F_s \frac{\partial }{\partial p^{\alpha}_s}, \quad 
\frac{\partial}{\partial {\bf q}^{\alpha}} \left( n({\bf q},t) 
\langle A {\bf G} \rangle_{p,\alpha} \right)=\sum^m_{s=1}
\frac{\partial}{\partial q^{\alpha}_s} \left( n({\bf q},t) 
\langle A G_s \rangle_{p,\alpha} \right) . \]

\section{Hydrodynamic Equation for Fractional Systems}

Let us consider the special cases of transport 
equation (\ref{EE}) for 
\[ A=M, \quad A=p^{\alpha}_s= MG_s, \quad A = 
\frac{{\bf p}^{2\alpha}}{2M} . \]
If we use $A=M$, then (\ref{EE}) gives
\be \label{EE1}
\frac{\partial}{\partial t} \tilde \rho_M({\bf q},t) 
+\frac{\partial}{\partial q^{\alpha}_s} \tilde \rho_M({\bf q},t) 
\langle G_s \rangle_{p,\alpha}({\bf q},t) =0 , \ee
where $\rho_M$ is mass density
\be \label{rM} \tilde \rho_M({\bf q},t)=M n({\bf q},t) . \ee
For $A=p^{\alpha}_s=MG_s$, we obtain
\be \label{EE2} 
\frac{\partial}{\partial t} \tilde \rho_M V_l({\bf q},t)+
\frac{\partial}{\partial q^{\alpha}_s} \tilde \rho_M  
\langle G_s G_l  \rangle_{p,\alpha} = \tilde \rho_M({\bf q},t) 
Z \langle F_l \rangle_{p,\alpha},  \ee
where we use Eqs. (\ref{Gs}), (\ref{rM}),  and the relation
\[ \frac{\partial G_l}{\partial p^{\alpha}_s}=M\delta_{ls} . \]
For $A ={{\bf p}^{2\alpha}}/{2M}$, we get 
\be \label{EE3} 
\frac{\partial}{\partial t} \tilde \rho_M({\bf q},t) 
\langle \frac{{\bf p}^{2 \alpha}}{2M^2} \rangle_{p,\alpha}+
\frac{\partial}{\partial q^{\alpha}_s} \tilde \rho_M ({\bf q},t) 
\langle \frac{1}{2}G^2_l G_s  \rangle_{p,\alpha} = 
\tilde \rho_M({\bf q},t) Z \langle 
F_s G_s \rangle_{p,\alpha} .  \ee
Here $\langle F_s G_s \rangle_{p,\alpha} $ is a local mean value.

Let us define the deviation of velocity from its mean value by
\be  C_s({\bf q},t) =G_s-V_s({\bf q},t)=
G_s-\langle G_s \rangle_{p,\alpha} . \ee
Substituting $G_s =V_s+C_s$ in the kinetic energy tensor 
$\langle G_s G_l \rangle_{p,\alpha}$, we get
\be \label{GG} \langle G_s G_l \rangle_{p,\alpha}=V_sV_s-
\langle C_s C_l \rangle_{p,\alpha} , \ee
where we use $\langle C_s \rangle_{p,\alpha}=0$. 
From Eq. (\ref{GG}), we have
\be \label{v2-1} 
\langle \frac{{\bf p}^{2\alpha}}{2M}\rangle_{p,\alpha}
=\frac{M{\bf V}^2}{2}+\frac{M{\bf C}^2}{2} . \ee

The tensor of internal stress 
\be \label{Psl} P_{sl}=\langle C_s C_l \rangle_{p,\alpha} , \ee
can be represented as the sum 
\[ P_{sl}({\bf q},t)=\delta_{sl}P({\bf q},t)+\pi_{sl}({\bf q},t) , 
\quad \pi_{ss}({\bf q},t)=0 , \]
where $\pi_{sl}({\bf q},t)$ is the tensor of viscous stress. 
Then 
\[ \frac{\rho_M}{2}\langle G^2_l G_s  \rangle_{p,\alpha}=
\frac{\rho_M}{2}\langle (V_l+C_l)^2 (V_s+C_s)  \rangle_{p,\alpha}=\]
\[ =\frac{\rho_M}{2}
\langle (V^2_l+C^2_l+2V_lC_l) (V_s+C_s)  \rangle_{p,\alpha} =\]
\be \label{v2-2} 
=\frac{\rho_M}{2}(V^2_s+\langle C^2_l \rangle_{p,\alpha}) V_l+
2V_lP_{ls}+\frac{\rho_M}{2} \langle C^2_l C_s  \rangle_{p,\alpha} . \ee

Substituting Eq. (\ref{Psl}) in Eq. (\ref{EE2}), and substituting 
Eqs. (\ref{v2-1}), (\ref{v2-2}) in Eq. (\ref{EE3}),
we get the equations for the functions
$\rho_M({\bf q},t)$, $V_s({\bf q},t)$, $E({\bf q},t)$.
From (\ref{EE1}), we get the continuity equation
for fractional systems
\cite{chaos,PRE05}:  
\be \label{HE1}
\frac{\partial}{\partial t} \tilde \rho_M 
+\frac{\partial}{\partial q^{\alpha}_s} \tilde \rho_M V_s =0 , \ee
which may be regarded as the equation of balance of 
"fractional matter". This matter can be described by fractional 
systems (\ref{H3}). 
In addition to $\rho_M$, the continuity equation (\ref{HE1}) 
includes the density of momentum $\rho_M {\bf V}$. 
To obtain the equation for the density of momentum,
we multiplied first fractional Bogoliubov equation 
by ${\bf p}^{\alpha}$, and use fractional integration 
over ${\bf p}$. Taking advantage of the assumption 
(\ref{ass}) and the boundary condition (\ref{bound}),
we get the equation for the components of 
the vector of density of momentum 
\be \label{HE2} \frac{\partial}{\partial t} \tilde \rho_M V_l+
\frac{\partial}{\partial q^{\alpha}_s} 
(\tilde \rho_M V_s V_l +P_{sl}) =f_l , \ee
where $f_l=f_l({\bf q},t)=Z \langle F_s \rangle_{p,\alpha}$. 

Finally, to write down the equation of balance of fractional
kinetic energy density, we multiplied first fractional 
Bogoliubov equation by ${\bf p}^{\alpha}$, and use fractional 
integrated with respect to ${\bf p}$. Taking advantage of the assumption 
(\ref{ass}) and the boundary condition (\ref{bound}),
we come to the equation for the density of kinetic energy of fractional systems:
\be \label{HE3}
\frac{\partial}{\partial t} \left(\frac{1}{2}\tilde \rho_M V^2+E\right)+
\frac{\partial}{\partial q^{\alpha}_s} 
\left( V_s \left[\frac{\tilde\rho_M V^2}{2} +E \right]
+P_{sl}V_l+Q_s\right)=f_s V_s , \ee
where 
\[ Q_s=\frac{\rho_M}{2} \langle C^2_l C_s  \rangle_{p,\alpha} . \]
Equations (\ref{HE1}), (\ref{HE2}), and (\ref{HE3}) are the 
hydrodynamic equations for fractional systems.
Obviously, the set of this five equations is not closed.
If we have $\pi_{sl}=0$ and $Q_s=0$, then these equations are
\be \label{HE10}
\frac{\partial}{\partial t} \tilde \rho_M 
+\frac{\partial}{\partial q^{\alpha}_s} \tilde \rho_M  
\langle G_s \rangle_{p,\alpha}  =0 . \ee
\be \label{HE20} \frac{\partial}{\partial t} \tilde \rho_M V_l+
\frac{\partial}{\partial q^{\alpha}_s} 
(\tilde \rho_M V_s V_l)=-\frac{\partial P}{\partial q^{\alpha}_l}+f_l , \ee
\be \label{HE30}
\frac{\partial}{\partial t} \left(\frac{1}{2}\tilde \rho_M V^2+E\right)+
\frac{\partial}{\partial q^{\alpha}_s} 
\left( V_s \left[\frac{\tilde\rho_M V^2}{2} +E+P \right] \right)=f_s V_s , \ee
and the set of equations is closed.

\section{Conclusion}

In this paper the hydrodynamic equations for fractional systems 
are derived.
In order to derive these equations, we use
the first Bogoliubov equation for fractional systems \cite{PRE05}.
Then we define the fractional generalization of the average values 
and the reduced distribution functions.
The Enskog equation for fractional systems is considered.
Hydrodynamic equations (\ref{HE1}), (\ref{HE2}), 
and (\ref{HE3}) can be considered as equations 
in the fractional space \cite{chaos,PRE05} or 
for systems with non-Gaussian statistics \cite{chaos,PRE05}.

Dissipative and non-Hamiltonian systems can have
stationary states of the Hamiltonian systems \cite{Tarpre}.
Classical dissipative systems can have
canonical Gibbs distribution as solutions of 
Liouville equations for the dissipative systems \cite{Tar-mplb,JPA05,AP05}.
Using the methods \cite{Tar-mplb,AP05} it is easy to find
solutions for the Bogoliubov equations for fractional systems.
Suggested Bogoliubov equation allows to formulate the dynamics 
for fractional generalization of quantum dissipative systems
by methods suggested in \cite{Tarpla1,Tarmsu,JPA04}.



\end{document}